# Turbulence in the Local Interstellar Medium and the *IBEX* Ribbon

E. J. Zirnstein[1], J. Giacalone[2], R. Kumar[1], D. J. McComas[1], M. A. Dayeh[3], J. Heerikhuisen[4]

[1]Department of Astrophysical Sciences, Princeton University, Princeton, NJ 08544, USA
(ejz@princeton.edu)
[2]Department of Planetary Sciences, University of Arizona, Tucson, AZ 85721, USA
[3]Southwest Research Institute, San Antonio, TX 78228, USA
[4]Department of Space Science, University of Alabama in Huntsville, Huntsville, AL 35899, USA

Abstract:

The effects of turbulence in the very local interstellar medium (VLISM) have been proposed by Giacalone & Jokipii (2015) to be important in determining the structure of the *Interstellar Boundary Explorer* (*IBEX*) ribbon via particle trapping by magnetic mirroring. We further explore this effect by simulating the motion of charged particles in a turbulent magnetic field superposed on a large-scale mean field, which we have considered to be either spatially-uniform or a mean field derived from a 3D MHD simulation. We find that the ribbon is not double-peaked, in contrast to Giacalone & Jokipii (2015). However, the magnetic mirror force still plays an important role in trapping particles. Furthermore, the ribbon's thickness is considerably larger if the large-scale mean field is draped around the heliosphere. *Voyager 1* observations in the VLISM show a turbulent field component that is stronger than previously thought, which we test in our simulation. We find that the inclusion of turbulent fluctuations at scales ≳100 au and power consistent with *Voyager 1* observations produces a ribbon whose large-scale structure is inconsistent with *IBEX* observations. However, restricting the fluctuations to ~10 au or smaller produces a smoother ribbon structure similar to *IBEX* observations. Different turbulence realizations produce different small-scale features (≲10°) in the ribbon, but its large-scale structure is robust if the maximum fluctuation size is ~10 au. This suggests that the magnetic field structure at scales ≲10 au is determined by the heliosphere's interaction with the VLISM and cannot entirely be represented by homogeneous interstellar turbulence.

## 1. Introduction

The supersonic solar wind (SW) plasma, consisting mostly of protons, electrons, and a few percent alpha particles, flows radially away from the Sun typically at speeds from ~300 to 800 km s$^{-1}$. The SW slows down to subsonic speeds at the termination shock (~100 au from the Sun), and the subsonic SW interacts with the partially-ionized interstellar gas in the very local interstellar medium (VLISM), forming the heliosphere with a tangential discontinuity (i.e., the heliopause) that separates the solar and interstellar plasma (e.g., Zank 1999; 2015). The neutral matter from the interstellar medium can cross the heliopause and travel into the inner heliosphere. Moreover, interstellar neutral atoms, mostly hydrogen and helium, can charge-exchange with SW ions, generating "pickup ions from neutral interstellar wind" (PINI) as well as energetic neutral atoms (ENAs), which can propagate large distances before ionizing.

The *Interstellar Boundary Explorer* (*IBEX*; McComas et al. 2009a) is an Earth-orbiting spacecraft with two single-pixel cameras (Funsten et al. 2009; Fuselier et al. 2009) that detect interstellar neutral atoms flowing into the heliosphere from the VLISM (e.g., Bzowski et al. 2017; Kubiak et al. 2016; McComas et al. 2015; Möbius et al. 2009; Park et al. 2016; Schwadron et al. 2016) and ENAs produced by charge-exchange in the outer heliosphere (e.g., McComas et al. 2009b, 2017, 2018b; Desai et al. 2019; Schwadron et al. 2018). Full-sky observations of neutral atoms provide a means to deduce the thermodynamic and structural properties of the outer heliosphere and its interaction with the VLISM (e.g., Bzowski et al. 2017; Zirnstein et al. 2017).

*IBEX* discovered an unpredicted enhancement of ENAs forming a nearly complete circle across the celestial sky (McComas et al. 2009b); these authors also showed that the ribbon may contain fine-scale structure. This "ribbon" of enhanced ENAs has been measured and studied ever since its discovery in 2009. The ribbon flux is strongly correlated with the local interstellar magnetic field (ISMF) draped around the heliosphere (Schwadron et al. 2009; Heerikhuisen & Pogorelov 2011; Pogorelov et al. 2011; Zirnstein et al. 2016), and it is believed to be formed from "secondary" ENAs produced by charge-exchange between "pickup ions from neutral solar wind" (PINS) and interstellar neutral atoms in the draped ISMF outside the heliopause (e.g., McComas et al. 2009b, 2017; Zirnstein et al. 2015a). Numerous models have furthered our understanding of the ribbon's origin from the secondary ENA mechanism (e.g., Heerikhuisen et al. 2010; Chalov et al. 2010; Gamayunov et al. 2010; Möbius et al. 2013; Schwadron & McComas 2013; Isenberg 2014, 2015; Giacalone & Jokipii 2015, hereafter GJ15; Zirnstein et al. 2018), focusing our attention on whether or not the parent PINS experience significant pitch angle scattering outside the heliopause (e.g., Florinski et al. 2010; 2016; Summerlin et al. 2014; Niemiec et al. 2016), whether the interstellar turbulence is important (GJ15; Gamayunov et al. 2017, 2019), how different extremes of pitch angle scattering affect the ribbon flux observed at 1 au (e.g., Zirnstein et al. 2018, 2019; Gamayunov et al. 2019), and how the ribbon's source distribution and the $\boldsymbol{B} \cdot \boldsymbol{r} = 0$ surface are related to the draping of the ISMF around the heliosphere (e.g., Grygorczuk et al. 2011; Strumik et al. 2011; Ratkiewicz et al. 2012; Isenberg et al. 2015; Zirnstein et al. 2015a, 2016).

One of the possible sources of the *IBEX* ribbon is its formation in the interstellar turbulence (GJ15). These authors showed that the propagation of PINS in the presence of homogeneous turbulence with a uniform mean magnetic field in the VLISM can produce a ribbon of width ~10° due to a magnetic mirror force that traps particles with high pitch angles near $\boldsymbol{B} \cdot \boldsymbol{r} = 0$. GJ15 also predict that this causes the ribbon emission to have a double-humped flux profile perpendicular to the ribbon; however, we have found this to be a numerical artifact of the method they used and is not really, as we discuss further below. But an important question remains: what does the ribbon observed at 1 au look like when there is interstellar turbulence and the ISMF is draped around the heliosphere? Moreover, recent *Voyager 1* observations of turbulence in the VLISM suggests that the level of turbulence at length scales <5 au is much larger that the estimated interstellar turbulence at this scale (Burlaga et al. 2018). Then the question arises: how does this increased power in the turbulent fluctuations affect the ribbon?

In this study, we implement the model presented by GJ15, namely the propagation of PINS in the presence of turbulence outside the heliopause. We extend the work of GJ15 by including a homogeneous magnetic turbulent component that is superposed on a mean field derived from a three-dimensional (3D), magnetohydrodynamic (MHD) simulation of the heliosphere. The 3D MHD simulation models the draping of the large-scale ISMF around the heliopause. We present the model assumptions in Section 2, present results of simulated ENA fluxes at 1 au in Section 3 and discuss their implications for the origin of the *IBEX* ribbon and turbulence in the VLISM in Section 4.

## 2. Model

### 2.1. Simulation of the Heliosphere

Similar to our earlier work on modeling the ribbon (e.g., Zirnstein et al. 2018, 2019), we utilize plasma and neutral results from a 3D, global simulation of the SW-VLISM interaction. The simulation iterates between MHD-plasma and kinetic (Boltzmann)-neutral modules, coupled by energy-dependent, charge-exchange source terms, to simulate the heliosphere (e.g., Pogorelov et al. 2008, 2009a; Heerikhuisen et al. 2009, 2013). The VLISM boundary conditions, applied at a radius of 1000 au from the Sun, were derived based on constraining the *IBEX* ribbon's position in the sky (Zirnstein et al. 2016), producing an ISMF magnitude of 2.93 µG and direction (227.28°, 34.62°) in ecliptic J2000 coordinates at the outer boundary of the simulation (1000 au). This field produces a draped magnetic field at the location of *Voyager 1* that is consistent with its measurements of the field magnitude and orientation. The interstellar neutral temperature (7500 K, assumed to be the same for ions), flow speed (25.4 km s$^{-1}$), and inflow direction (255.7°, 5.1°) are derived from *IBEX* observations of interstellar neutral atoms (McComas et al. 2015). With these boundary conditions and interstellar plasma density of 0.09 cm$^{-3}$ and neutral hydrogen density of 0.154 cm$^{-3}$, the simulation is also consistent with the (1) distance to the heliopause in the *Voyager 1* (Stone et al. 2013; Gurnett et al. 2013) and *Voyager 2* directions[1] (~120 au from the Sun), (2) the correlation between the interstellar neutral hydrogen deflection plane (Lallement et al. 2010) and the plane formed by the ISMF field direction (***B***) and the VLISM inflow direction (***V***), i.e., ***B-V*** plane (Zirnstein et al. 2016), and (3) the interstellar neutral hydrogen density at the termination shock near the VLISM inflow direction (~0.1 cm$^{-3}$; Bzowski et al. 2009).

The SW boundary conditions at 1 au are the same as those used in our previous work: SW plasma density is 5.74 cm$^{-3}$, plasma temperature is 51,100 K, flow speed is 450 km s$^{-1}$, and magnetic field radial component in the Parker spiral is 37.5 µG. The SW values are advected to the simulation's inner boundary (10 au) by adiabatic expansion. The SW boundary conditions are assumed to be time and latitude-independent. Moreover, to avoid the artificial presence of a flat current sheet and spurious magnetic reconnection at the heliopause that may arise due to numerical dissipation, SW magnetic field in this study is unipolar.

### 2.2. The Neutralized SW

We create a neutralized SW distribution as the source of PINS outside the heliopause following Swaczyna et al. (2016b) and Zirnstein et al. (2019). We utilize results from a model of the SW speed and density derived from interplanetary scintillation (IPS) observations as a function of heliographic latitude (Sokół et al. 2015). We average SW speed and density at 1 au ($u_{SW,0}$ and $n_{SW,0}$, respectively) over time from 2000 through 2009, which accounts for (1) the predicted ~4 to 9 yr delay between SW observations at 1 au and *IBEX* ribbon observations at 1 au (Zirnstein et al. 2015b) and (2) the time-averaged, first 5 years of *IBEX* observations from

---

[1] https://voyager.jpl.nasa.gov/news/details.php?article_id=112

2009 to 2013 (McComas et al. 2014), which we compare to later in the paper. Thus, the neutralized SW differential flux,

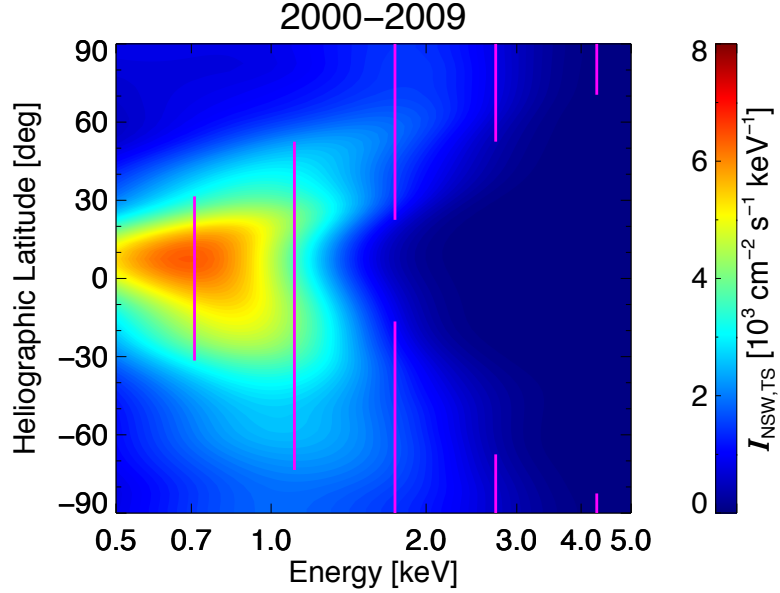

**Figure 1.** Neutral SW flux at the termination shock, $I_{NSW,TS}$, as a function of energy and heliographic latitude. We show data time-averaged from 2000 through 2009, which approximately corresponds to the observation time of IBEX from 2009-2013. The magnetic lines show the central energies of the *IBEX-Hi* energy channels, with ranges in latitude that cover fluxes greater than half of the maximum at that energy.

$I_{NSW}$, at the radial distance of the termination shock, $r_{TS}$ (which we assume to be 100 au), is given by (e.g., Swaczyna et al. 2016b)

$$I_{NSW}(r_{TS}, v, \theta) =$$
$$\frac{1}{M}\Sigma_i \int_{r_0}^{r_{TS}} [n_{SW,0}(\theta,i) u_{SW,0}(\theta,i)] \left(\frac{r_0}{r_{TS}}\right)^2 [n_H \sigma_{ex}(v)] e^{-n_H \sigma_{ex}(v) r} N(u_{SW,i}(r), \delta v|v) \frac{1}{m_p v} dr, \quad (1)$$

where $v$ is the neutral SW speed, $\theta$ is the heliographic latitude, $i$ is the Carrington rotation number summed over 1958 through 2091 (total $M$ = 134), $r_0$ = 1 au, $n_H$ = 0.09 cm$^{-3}$ is the interstellar neutral hydrogen density in the supersonic SW (e.g., Bzowski et al. 2009), $\sigma_{ex}$ is the energy-dependent, charge-exchange cross section (Lindsay & Stebbings 2005), $r$ is the radial distance from the Sun, $m_p$ is the proton mass, and $N$ is a Gaussian speed distribution with mean speed $u_{SW,i}(r)$ and thermal speed $\delta v$ = 100 km s$^{-1}$ which smooths the SW in speed (Swaczyna et al. 2016b). The SW slows down farther from the Sun due to mass-loading from PINI. This effect is approximated by (e.g., Lee et al. 2009)

$$u_{SW}(r) = u_{SW,0} \left[1 - \left(1 - \frac{1}{2}\frac{\gamma-1}{2\gamma-1}\right)\frac{r}{\lambda_{ml}}\right],$$

$$\lambda_{ml} = \left[n_H \sigma_{ex} + \frac{\nu_H n_H + 4\nu_{He} n_{He}}{n_{SW,0} u_{SW,0}}\right]^{-1}, \quad (2)$$

where $\gamma$ = 5/3 is the adiabatic index, $n_{He}$ = 0.015 cm$^{-3}$ is the interstellar neutral helium in the supersonic SW (Gloeckler et al. 2004), and $\nu_H = \nu_{He} = 10^{-7}$ s$^{-1}$ are the photo-ionization rates for hydrogen and helium atoms, respectively, at 1 au (Bzowski et al. 2013; Sokół et al. 2019). Integrating Equation (1) to $r_{TS}$ gives the SW differential flux at the SW termination shock, averaged from 2000 through 2009, which is plotted in Figure 1 (for more details, see Swaczyna et al. 2016b; Zirnstein et al. 2019).

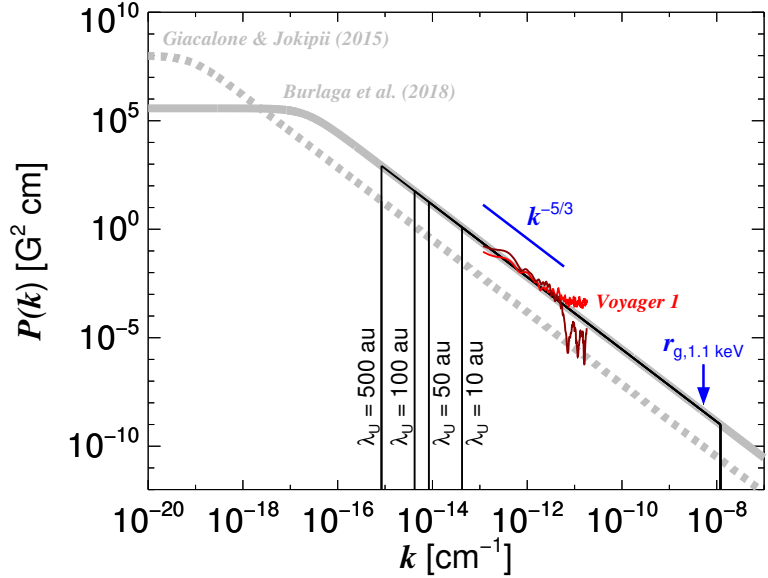

**Figure 2.** Kolmogorov power-law spectra with different parameters used in this study. Theoretical spectra (solid/dashed gray and black curves) are calculated using Equation (9). The power spectrum similar to that used by GJ15 is shown as the dashed gray curve ($L_C$ = 4 pc, $\sigma_C$ = 4 μG), and the power spectrum similar to Burlaga et al. (2018) required to match the *Voyager 1* observations is shown as the solid gray curve ($L_C$ = 0.015 pc, $\sigma_C$ = 4 μG). We show *Voyager 1* data from Burlaga et al. (2018) for time periods 2013.3593-2014.6373 (dark red) and 2015.3987-2016.6759 (light red). The scale of a proton gyroradius ($r_g$) with energy 1.1 keV, equivalent to the central energy of the *IBEX-Hi* energy passband 3 (Funsten et al. 2009), in a 4 μG magnetic field is shown by the blue arrow. We simulate the ribbon in turbulence spectral cutoff scales (wavelength upper limit, $\lambda_U$), with a minimum wavelength of ~$0.5 \times r_{g,1.1keV}$.

To compute the neutral SW flux outside the heliopause, we account for (1) the $r^{-2}$ expansion of the neutral SW and (2) the loss by charge-exchange in the OHS. Thus, the neutral SW differential flux at distance $r$ from the Sun (here, $r > r_{HP}$) is

$$I_{NSW}(r,v) = I_{NSW}(r_{TS}, v, \theta) \left(\frac{r_{TS}}{r}\right)^2 e^{-\int_{r_{HP}}^{r} n_p(r')\sigma_{ex}dr'}. \quad (3)$$

### 2.3. Turbulent Magnetic Field

Following Giacalone & Jokipii (1999), we construct a 3D turbulent magnetic field component by superimposing a large number of shear Alfvén waves of random polarizations, phases, and wavevector directions. For a large enough number of wave modes, this method yields an isotropic and spatially homogeneous turbulent field (Batchelor 1960). The random component of the magnetic field, $\delta B$, is (see also Giacalone & Jokipii 1999 for more details)

$$\delta \boldsymbol{B}(x,y,z) = \sum_{n=1}^{N_m} A(k_n) \hat{\boldsymbol{\xi}}_n \exp(ik_n z'_n + i\beta_n), \quad (4)$$

$$\hat{\boldsymbol{\xi}}_n = \cos \alpha_n \, \boldsymbol{x}'_n + i\sin \alpha_n \, \boldsymbol{y}'_n, \quad (5)$$

$$\begin{pmatrix} x' \\ y' \\ z' \end{pmatrix} = \begin{pmatrix} \cos\theta_n \cos\varphi_n & \cos\theta_n \sin\varphi_n & -\sin\theta_n \\ -\sin\varphi_n & \cos\varphi_n & 0 \\ \sin\theta_n \cos\varphi_n & \sin\theta_n \sin\varphi_n & \cos\theta_n \end{pmatrix} \begin{pmatrix} x \\ y \\ z \end{pmatrix}, \quad (6)$$

where $n$ is the wave mode number out of total $N_m$, $k_n$ is the wavenumber, $\alpha_n$ is the polarization, $\beta_n$ is the phase, and the propagation direction of each wave mode is represented by angles $\theta_n$ and $\varphi_n$. For isotropic turbulence, $\alpha_n$, $\beta_n$, $\theta_n$, and $\varphi_n$ are selected randomly from a uniform distribution in the following ranges: $0 < \alpha_n < 2\pi$, $0 < \beta_n < 2\pi$, $-1 < \cos(\theta_n) < 1$, and $0 < \varphi_n < 2\pi$. For the results presented in this study we set $N_m = 500$.

The wave amplitude $A(k_n)$ is represented by a Kolmogorov spectrum, such that

$$A^2(k_n) = \sigma^2 G(k_n) / \left[ \sum_{n=1}^{N_m} G(k_n) \right], \quad (7)$$

$$G(k_n) = \frac{4\pi k_n^2 \Delta k_n}{1 + (k_n L_C)^\gamma}, \quad (8)$$

where $\gamma = 11/3$ for 3D Kolmogorov spectrum and $\sigma^2$ is the wave variance. Similar to Giacalone & Jokipii (1999), the logarithm of wavenumbers are equally spaced.

Following GJ15, for each simulation presented in this study, we construct a turbulence spectrum with an upper limit scale of 20 pc and lower limit scale of $0.5 \times r_g$, where $r_g$ is the gyroradius of a 400 km s$^{-1}$ proton in a 4 µG magnetic field. The total magnetic field is given by $\boldsymbol{B}(x,y,z) = \boldsymbol{B}_0(x,y,z) + \delta\boldsymbol{B}(x,y,z)$, where $\boldsymbol{B}_0$ is the mean magnetic field. For the majority of the results presented in this study, we set $\boldsymbol{B}_0$ equal to the magnetic field from our 3D MHD simulation of the heliosphere interaction with the VLISM ($\boldsymbol{B}_{MHD}$; e.g., Pogorelov et al. 2009b; Heerikhuisen et al. 2014; Zirnstein et al. 2016). Since we desire the total mean field to be dominated at large scales ($\gtrsim$100 au) by the draped MHD field, when summing over wave modes at each iteration of the particle stepping, we integrate up to a cutoff scale similar to or smaller than the scale size of the heliosphere.

An analytic expression for the 1D Kolmogorov turbulent power spectrum, which we use to plot the power spectrum in Figure 2, is given by (e.g., Appendix B in Giacalone & Jokipii 1999)

$$P(k) = \frac{\frac{\sigma^2}{1+(kL_C)^{5/3}}}{\left[ \int_0^\infty \frac{dk}{1+(kL_C)^{5/3}} \right]} = \frac{5\sin(3\pi/5)}{3\pi} \frac{\sigma^2 L_C}{1+(kL_C)^{5/3}}. \quad (9)$$

We note that the expression used by Burlaga et al. (2015, 2018) is different than our Equation (9) by a factor of 2 because they integrate over wavenumbers from $-\infty$ to $\infty$ and replace $k$ with $|k|$ in Equation (9). We use the form of Equation (9) presented here to clarify that $k$ is the magnitude of the wavevector and must be positive; therefore, $k$ is integrated from 0 to $\infty$.

**2.4. Solving Particle Motion**

Following GJ15, we numerically integrate the equation of motion for protons in a turbulent magnetic field outside the heliopause. However, in this study we define the heliopause surface using the MHD simulation. The motion of protons in a magnetic field is governed by the Lorentz force equation (written in non-relativistic form, in cgs units),

$$\frac{d\boldsymbol{v}_p}{dt} = \frac{q}{m_p c} \boldsymbol{v}_p \times \boldsymbol{B}, \quad (10)$$

where $v_p$ is the PINS velocity (note that $|v_p| = v$, the speed used in Equations 3 and 12), $q$ is the proton charge, $m_p$ is the proton mass, $c$ is the speed of light, and $B$ is the total (mean $B_0$ + turbulent $\delta B$) magnetic field at the particle's instantaneous position. We integrate Equation (10) using the Bulirsch-Stoer method (Press et al. 2002). We use adaptive stepping to track the error such that it does not exceed a tolerance limit during each step. In the adaptive Bulirsch-Stoer method, the magnetic field is calculated at multiple discrete locations during each time step in order to render a desired accuracy of a particle's trajectory. For the majority of the results presented in this study, the tolerance is set such that over a typical charge-exchange lifetime of a proton outside the heliopause (~2 yr), the accumulated error in particle energy is ≲5%. For results compared to *IBEX* data (Figure 6), we decrease the tolerance level such that the accumulated error in particle energy is <2% over a typical charge-exchange lifetime.

There is a key difference in the methods used in this study to simulate the *IBEX* ribbon compared to, e.g., GJ15. While those authors initialized millions of protons extracted from a neutral SW distribution outside the heliopause and propagated the particles forward in time until they charge-exchange into secondary ENAs, in this study we propagate particles backwards in time, similar to the methodology described by Zirnstein et al. (2018), which we summarize here. We step outward from the Sun along an *IBEX* line-of-sight in small, discrete intervals $\Delta r$ (the distance interval for the ENA integration, see Equation 12). At each step outside the heliopause, we begin an integration of Equation (10) backwards in time starting from the current position $r$, which is the position at which a secondary ENA with desired speed and propagation direction would intersect *IBEX*'s line-of-sight. We integrate Equation (10) from $t = 0$ to $t_{max}$, where $t_{max} > \tau_{ex} = (n_H \sigma_{ex} v)^{-1}$ and $\tau_{ex}$ is the charge-exchange mean free lifetime. For this study we set $t_{max} = 2\tau_{ex}$.

During each time step $\Delta t$ of the integration of Equation (10) from $t = 0$ to $t_{max}$ (where we set $\Delta t$ equal to the inverse gyrofrequency $\Omega_g^{-1} = m_p/(qB)$), we record the current velocity vector of the proton $v'$ at current position $r'$ and calculate the local production of PINS (at $r'$ with velocity $v'$) by charge-exchange using the neutral SW distribution from Section 2.2. The contribution of these PINS to the secondary ENA flux back at the start position ($r$, $t = 0$) (see Equations 13 and 14) is weighted by the probability for these PINS to survive traveling back to ($r$, $t = 0$). This weighting is given by (Zirnstein et al. 2015b)

$$W(t) = \frac{\exp(-t/\tau_{ex})}{\int_0^{t_{max}} \exp(-t'/\tau_{ex}) dt'}, \qquad (11)$$

which we normalize such that the integral of $W(t)$ over $t = 0$ to $t_{max}$ equals 1. The form for $W(t) \propto \exp(-t/\tau_{ex})$ represents the charge-exchange lifetime of the particle, which drops exponentially over time. We calculate $\tau_{ex}$ once at $t = 0$ for computational simplicity. This is a reasonable assumption because most PINS do not travel far during their lifetime (<10 au), which is much smaller than the scale over which the interstellar neutral density changes significantly (e.g., Heerikhuisen et al. 2014). The application of Equation (11) implies that the farther away from ($r$, $t = 0$) that a PINS originates, the less likely it could contribute to secondary ENAs created at ($r$, $t = 0$).

After computing the contribution of PINS to the secondary ENA flux at ($r$, $t = 0$), for distance interval $\Delta r$, we take another step outward along the *IBEX* line-of-sight and repeat the process described above. We repeat this process until we reach the outer boundary of the simulation, which we set to 600 au from the Sun.

The secondary ENA differential flux equation from PINS outside the heliopause is given by

$$J_{\text{ENA}}(\Omega, v) = \frac{v^2}{m_\text{p}} \int_{r_{\text{HP}}}^{r_{\text{OB}}} f_0(\boldsymbol{r}, v) n_\text{H}(\boldsymbol{r}) \sigma_{\text{ex}} P(\boldsymbol{r}, v) \text{d}r, \qquad (12)$$

where $\Omega$ is the *IBEX* line-of-sight direction, $f_0$ is the PINS distribution, $r_{\text{HP}}$ is the distance to the heliopause in direction $\Omega$ (extracted from our MHD simulation) and $r_{\text{OB}} = 600$ au is the outer boundary. The PINS distribution is given by

$$f_0(\boldsymbol{r}, v) = \int_0^{t_{\max}} W(t) [S_{\text{NSW}}(\boldsymbol{r}, v) \tau_{\text{ex}}] \text{d}t, \qquad (13)$$

where $S_{\text{NSW}}$ is the PINS source function produced by charge-exchange of the neutral SW outside the heliopause. We take into account the fact that the neutral SW distribution has a finite, though small, thermal spread transverse to the radial direction. Based on typical SW conditions, the transverse temperature of the neutral SW at the termination shock, $T_{t,\text{TS}}$, is approximately $\sim 5 \times 10^3$ K, which drops off as $r^{-2}$ farther from the Sun due to expansion (Florinski & Heerikhuisen 2017). Therefore, the PINS source function is given by

$$S_{\text{NSW}}(\boldsymbol{r}, v) = \frac{I_{\text{NSW}}(r,v) m_\text{p}}{2\pi} n_\text{p}(r) \sigma_{\text{ex}} v \left[ \frac{1}{\pi \delta v_\text{t}^2(r)} e^{-\left(\frac{v_\text{t}}{\delta v_\text{t}(r)}\right)^2} H\left(90° - \cos^{-1}(\hat{\boldsymbol{v}}_\text{p} \cdot \hat{\boldsymbol{r}})\right) \right],$$

$$v_\text{t} = v \sin\left(\cos^{-1}(\hat{\boldsymbol{v}}_\text{p} \cdot \hat{\boldsymbol{r}})\right),$$

$$\delta v_\text{t}(r) = \sqrt{\frac{2 k_\text{B} T_{t,\text{TS}}}{m_\text{p}} \left(\frac{r_{\text{TS}}}{r}\right)^2}, \qquad (14)$$

where the last term in square brackets for $S_{\text{NSW}}$ represents the transverse component of the neutral SW distribution. We apply the Heaviside step function (*H*) because PINS can only be created in the same hemisphere as the radially-propagating neutral SW.

**Table 1.** Turbulence spectrum parameters for each model case presented in this study.

| Parameter: | Outer Correlation Scale, $L_\text{C}$ | Variance RMS at $L_\text{C}$, $\sigma_\text{C}$ [a] | Spectrum Wavelength Upper Limit, $\lambda_\text{U}$ | Variance RMS at $\lambda_\text{U}$, $\sigma_\text{U}$ | Spectrum Wavelength Lower Limit, $\lambda_\text{L}$ |
|---|---|---|---|---|---|
| Units: | pc (au) | µG | au | µG | au |
| Case 1 | 0.015 (3.1×10³) | 4 | 500 | 1.0 | 3.5×10⁻⁵ |
| Case 2 | 0.015 (3.1×10³) | 4 | 100 | 0.60 | 3.5×10⁻⁵ |
| Case 3 | 0.015 (3.1×10³) | 4 | 50 | 0.48 | 3.5×10⁻⁵ |
| Case 4 | 0.015 (3.1×10³) | 4 | 10 | 0.28 | 3.5×10⁻⁵ |

[a] RMS = root mean square

We also include the survival probability for ENAs, $P(\boldsymbol{r},v)$, traveling from their point of creation to the SW termination shock in Equation (12). We exclude the survival probability inside the termination shock to make our results directly comparable with survival probability-corrected *IBEX* data (e.g., McComas et al. 2017). Unlike our previous work, we do not integrate Equation (12) over the *IBEX-Hi* energy ranges, due to the computational cost of simulating the propagation of many particles at different energies. However, in Figure 6 where we compare to *IBEX* data, we include the *IBEX-Hi* angular collimator response (Funsten et al. 2009) in the model results.

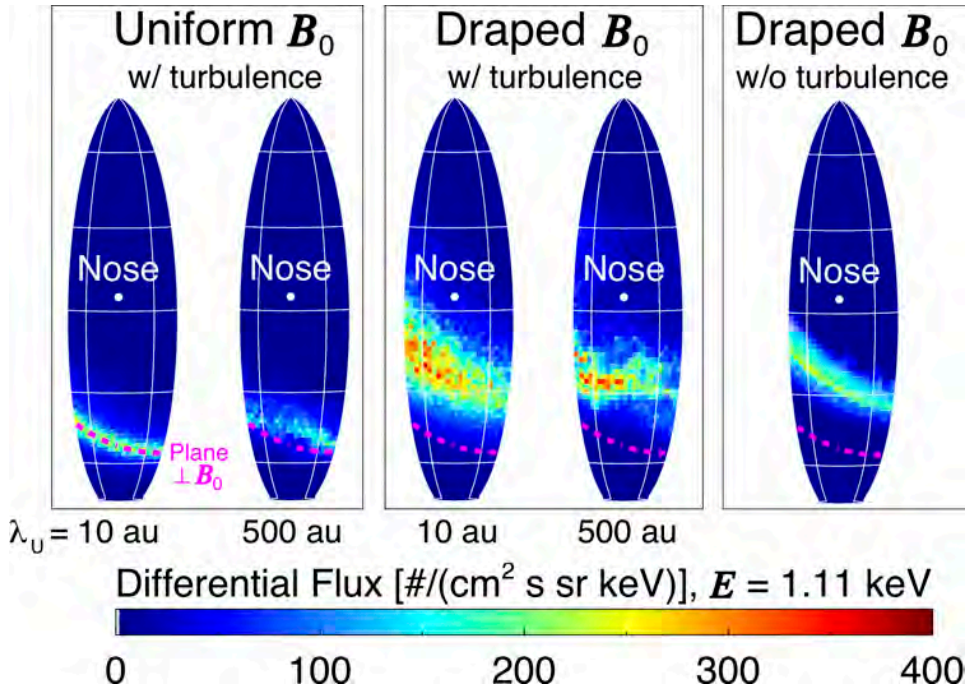

**Figure 3.** Model ribbon partial-sky maps at 1.1 keV for Cases 1 ($\lambda_U$ = 10 au) and 4 ($\lambda_U$ = 500 au) in Table 1, with turbulence superposed on a uniform mean field $B_0$ (left box) or a draped mean field derived from the MHD simulation ($B_0 = B_{MHD}$) (middle box). We also show the case where there is no turbulence (right box). The plane perpendicular to the (uniform) $B_0$ direction is shown as the dashed magenta curve, which is the same in each panel.

## 3. Results

In this section we present results from simulating ribbon ENA fluxes at 1 au from a source of PINS propagating in turbulence outside the heliopause. We also present results for different turbulence parameters (Table 1).

### 3.1. Uniform versus Draped Mean ISMF, $B_0$

First, we simulate the ribbon in a uniform mean ISMF with a turbulent component. We assume that the mean field, $B_0$, is directed towards (227.28°, 34.62°) in ecliptic J2000 coordinates, which is the pristine ISMF direction far from the heliosphere derived by Zirnstein et al. (2016), and set $|B_0|$ = 4 μG. Note that we choose $|B_0|$ = 4 μG (and not 2.93 μG, which was derived as the magnitude of the pristine/unaffected ISMF magnitude far from the heliosphere) in order to simulate the effects of a stronger, compressed mean field, similar to what happens to the ISMF draped around the heliopause, and (2) to emulate a magnitude similar to that used by GJ15. We also show results for the case of a self-consistently compressed and draped mean ISMF, $B_0 = B_{MHD}$, where $B_{MHD}$ is directly taken from our MHD simulation of the heliosphere using the best-fit parameters derived by Zirnstein et al. (2016). Results comparing these two cases are shown in Figure 3, alongside the case without turbulence.

As shown in Figure 3, the uniform mean ISMF assumption produces a significantly narrower ribbon compared to the case where the ISMF drapes around the heliosphere. The draping bends the $B_0 \cdot r = 0$ surface and naturally widens the angular portion of the sky that produces ENAs visible at 1 au (e.g., Pogorelov et al. 2011; Zirnstein et al. 2015a, 2016, 2019). The ribbon in the uniform mean ISMF case is also narrower than the uniform mean ISMF case presented by GJ15, likely because we limit the size of large-scale modes to <500 au. We show results for $\lambda_U = 10$ and 500 au (see Figure 2 and Table 1). For small $\lambda_U$, the uniform field case produces a ribbon that follows a great circle in the sky. For large $\lambda_U$, the turbulence power at large scales changes the mean field direction and causes the ribbon to meander across the sky, away from the great circle. The draped mean field case shows the effect of large-scale turbulence ($\lambda_U = 500$ au) on the ribbon's position and structure in the sky. Note also that turbulence at both scales widens the ribbon (compared to the right panel in Figure 3), since it allows particles at a broader range of pitch angles to have directions preferentially-aligned with *IBEX*'s line-of-sight.

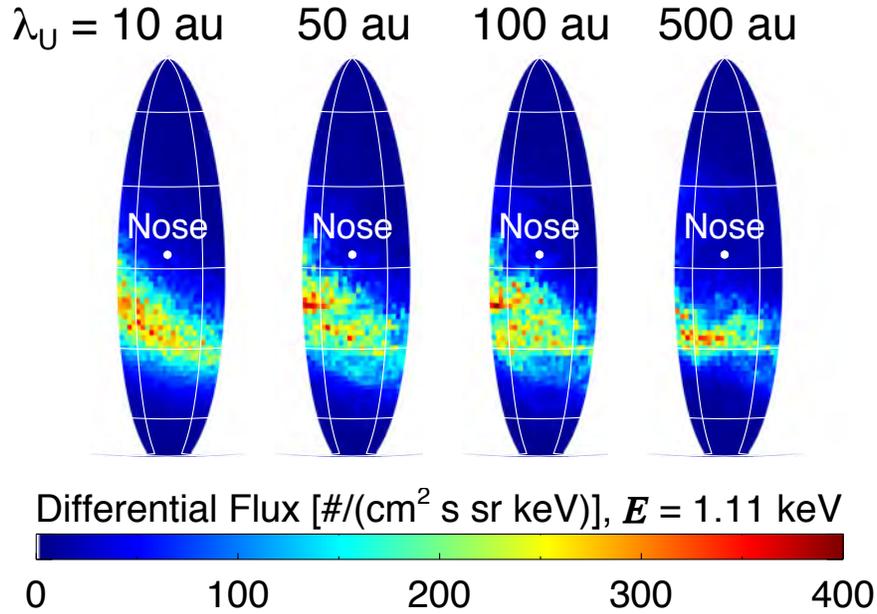

**Figure 4.** Model ribbon partial-sky maps at 1.1 keV for different spectrum upper limits, $\lambda_U$ (Cases 1-4 in Table 1) and a draped mean field $B_0$. Each case uses a turbulence spectrum that is normalized to *Voyager 1* power spectral density observations shown in Figure 2. We show results for a single turbulence realization.

### 3.2. Maximum Turbulence Scale

In this section we show how the ribbon changes when the maximum fluctuation scale size, $\lambda_U$, is varied. By varying $\lambda_U$, we include or exclude large-scale fluctuations and effectively change the (1) level of magnetic mirroring of particles and (2) the mean field direction. Figure 4 presents results where we varied $\lambda_U$ between 10 and 500 au, for a single turbulence realization in a draped mean ISMF. Note that "realization" refers to a unique set of random polarizations, phases, and propagation directions of waves in the turbulent magnetic field component. For small $\lambda_U$, the full-width at half maximum of the ribbon is approximately ~20° and the ribbon structure fluctuates on small scales (<10°), reminiscent of fine-scale structure possibly seen in the first *IBEX* maps (McComas et al. 2009b). As $\lambda_U$ increases, the ribbon structure becomes more distorted and chaotic at larger scales due to the inclusion of larger fluctuations in the magnetic field with larger amplitudes. For $\lambda_U = 500$ au, the peak of the ribbon meanders around the sky and is not consistent with the position or structure of the *IBEX* ribbon.

The results presented in Figure 4 are only for one unique realization of the turbulence and the ribbon structure may vary for different realizations, where each realization is constructed from a different set of random polarizations, phases, and wavevector directions. The dependence of the ribbon structure on the turbulence realization is demonstrated in more detail in Figure 5, which shows the simulated ribbon for three different turbulence realizations for $\lambda_U = 50$ au and $\lambda_U = 500$ au. For the smaller $\lambda_U$ case, while the different realizations change the small-scale structure and intensities, the large-scale structure remains roughly consistent. On the other hand, for $\lambda_U$ = 500 au, the ribbon at large scales changes drastically. In the first case, the ribbon appears almost parallel to the latitude = -30° line. For the second case, the ribbon is highly inclined with respect to the ecliptic plane. The large differences are due to the change in the large-scale fluctuations, which change the direction of the draped mean ISMF and alter the directions in the sky that the large-scale ribbon appears.

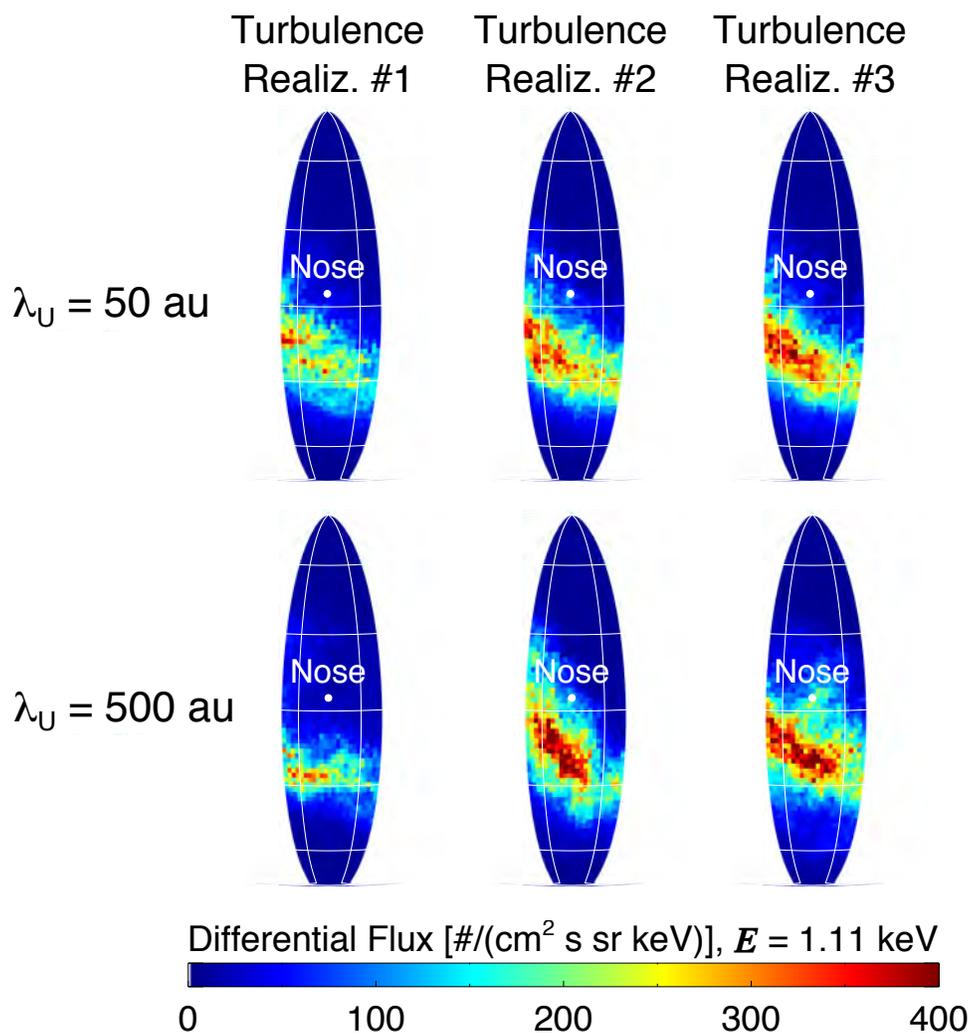

**Figure 5.** Model ribbon partial-sky maps at 1.1 keV for $\lambda_U = 50$ and 500 au, from three different turbulence spectrum realizations. Note that "Realization #1" is the same as Figure 4.

The results in Figures 4 and 5, when compared to multi-year averaged *IBEX* observations (see Figure 6), strongly suggest that the draped mean ISMF can be significantly altered by strong, large-scale turbulence in the VLISM, which affects the large-scale structure of the observed *IBEX* ribbon as shown in Figure 5 for $\lambda_U$ = 500 au. A comparison of our simulation for small $\lambda_U$ (e.g., $\lambda_U$ = 10 au) with multi-year averaged *IBEX* observations is shown in Figure 6.

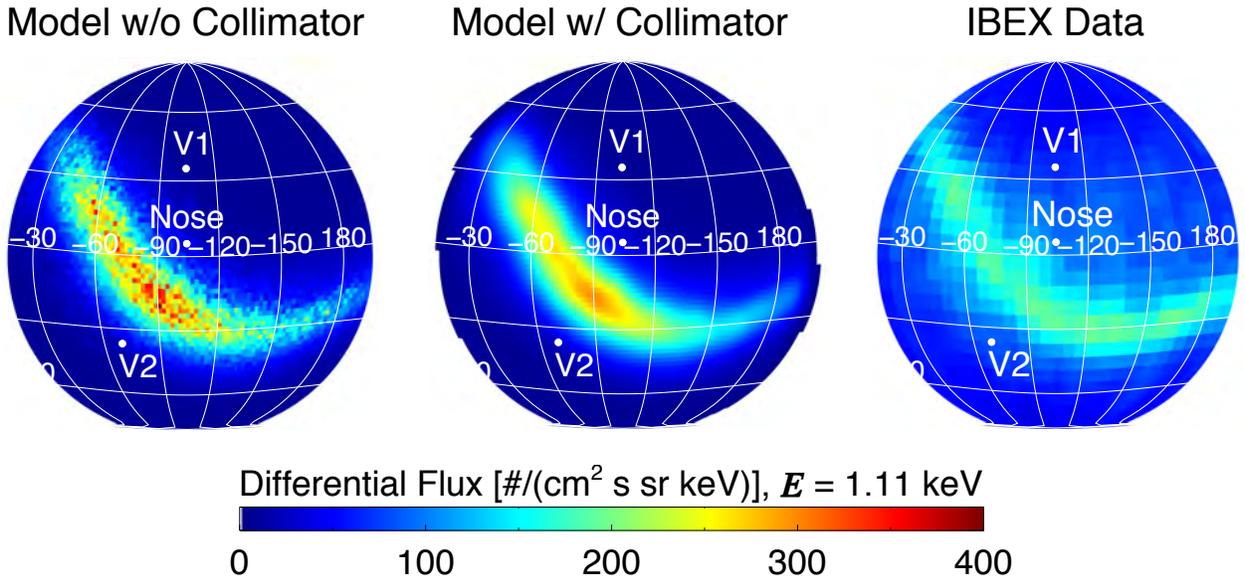

**Figure 6.** Model ribbon partial-sky maps at 1.1 keV for $\lambda_U$ = 10 au and Realization #1. For this result, we decreased the mover tolerance such that the error in energy of a 1.1 keV particle over a typical charge-exchange life time (~2 yr) is <2%. We show the model without the *IBEX* instrument angular collimator response (left), with the collimator response (middle), and 5 yr time-averaged *IBEX* data (right) from McComas et al. (2014). Note that the *IBEX* data also includes the globally distributed flux.

In Figure 6, we complete the simulation for the forward hemisphere of the sky (centered near the nose direction) for the case $\lambda_U$ = 10 au and Realization #1. We also show the simulated sky map after implementing the angular smoothing effects of the *IBEX* collimator (Funsten et al. 2009). The model compares reasonably well to the observations, in particular the width and position of the ribbon. Note, however, that we show the total intensity of *IBEX* observations, which includes the globally distributed fluxes emanating from the inner heliosheath that are not included in our simulation. While it appears that our simulation overestimates the observed intensities, it is possible a different realization of the turbulent field would yield a lower overall intensity in the ribbon (see Figure 5). Our simulation also does not directly account for any time-dependence in the ribbon source, and we do not integrate over *IBEX*'s energy response function, both of which may affect the simulated ribbon intensity.

### 3.3. Particle Pitch Angle Distribution in Turbulence

The results presented in this study show that the level of turbulence observed by *Voyager 1* and the maximum scale of the fluctuations significantly affects the propagation of particles outside the heliopause and the ability for them to produce ENAs that are visible by *IBEX*. In this section we further investigate the distribution of particles in the turbulence configurations presented so far.

Here, we perform a simple test particle simulation using the same Bulirsch-Stoer algorithm and turbulence realizations used in the results presented in Sections 3.1 and 3.2. We define a uniform mean field $\boldsymbol{B}_0$ to be aligned with the z-axis with magnitude $|\boldsymbol{B}_0| = 4$ µG and superpose the turbulent field component $\delta\boldsymbol{B}$ from the $\lambda_U = 10$ au and $\lambda_U = 500$ au cases shown in Figure 4. We follow ~50,000 particles with radial speed 400 km s$^{-1}$. They are initially randomly-distributed in a spherical grid with coordinates $(r, \theta, \varphi)$ limited to the following ranges: 100 au $< r <$ 1000 au, - $\sin(30°) < \cos(\theta) < \sin(30°)$, $-30° < \varphi < 30°$. Note that we only release particles in an angular window with opening angle of 60° centered on the *x*-axis to improve counting statistics near the radial direction perpendicular to the mean field, $\boldsymbol{B}_0 \cdot \boldsymbol{r} = 0$ (or latitude $\theta = 0°$). Then, as the particles propagate through space, we bin their pitch angle cosine, $\mu = \cos(\boldsymbol{v}_p \cdot \boldsymbol{B}/|\boldsymbol{B}|)$, and position every inverse gyrofrequency $\Omega_g^{-1} = m_p c/(qB_0)$. The lifetime of each particle, *t*, is randomly selected from an exponential distribution, $\exp(-t/\tau)$, with a 1/e mean lifetime $\tau = 500,000$ gyroperiods (1 gyroperiod = $2\pi\Omega_g^{-1}$). In a 4 µG magnetic field, 500,000 gyroperiods corresponds to approximately 2.5 yrs. The 1/e charge-exchange lifetime of a 1.1 keV proton outside the heliopause, assuming an interstellar neutral hydrogen density of 0.15 cm$^{-3}$, is also approximately 2.5 yrs.

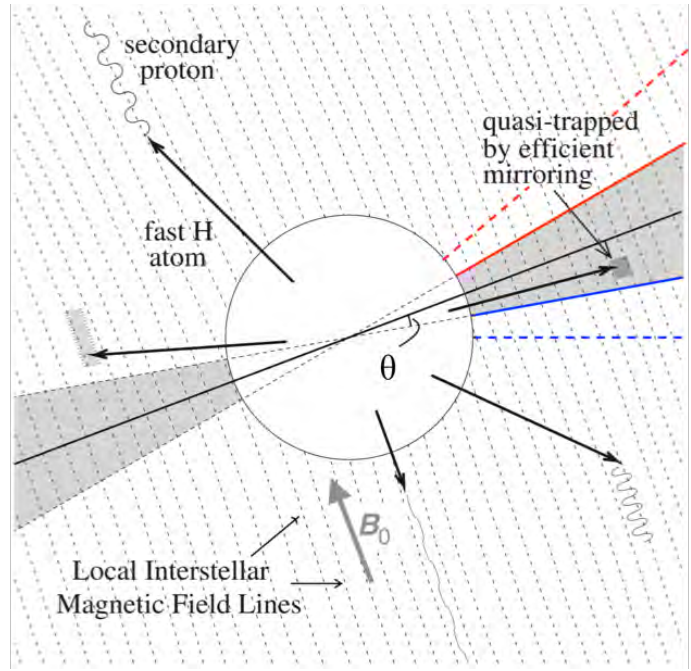

**Figure 7.** Illustration of the test particle system used to compute particle pitch angle distributions in Section 3.3. The directions in the sky from which the pitch angle distributions are extracted from are shown as the solid black ($\theta = 0°$), solid red and blue ($\theta = +10°$ and $-10°$) and dashed red and blue lines ($\theta = +20°$ and $-20°$). Adapted from GJ15.

Figures 8 and 9 show the particle pitch angle distributions for the same turbulence realizations shown in Figure 4 when $\lambda_U$ = 10 au and $\lambda_U$ = 500 au. Note that we bin particles for any radial distance *r*, but limited to $|\varphi|$ < 5°, into different latitude (θ) bins. At θ = 0°, which is directed toward $\boldsymbol{B}_0 \cdot \boldsymbol{r}$ = 0 (to mimic where the center of the ribbon peak would be located), the pitch angle distribution shows a single peak at μ = 0. This is in contrast to the results presented by GJ15, who showed a double-peaked pitch angle distribution in the center of the ribbon (see below). The distribution becomes broader for $\lambda_U$ = 500 au as particles experience a stronger mirror force from the larger fluctuations (and their larger amplitudes), causing the distribution to spread over a larger region of pitch angles around 90°. This effectively creates a wider ribbon, or even a change in the ribbon's position in the sky (see Figures 3 and 5).

Regarding the existence of the double-humped pitch angle distribution (and double-humped ribbon): we have studied this discrepancy carefully and have found that the double-humped feature is an artifact of the numerical method used by GJ15 and is not real. GJ15 made an unreported modification to the Bulirsch-Stoer numerical integration method which had the effect of forcing the magnetic field to be constant over a very small, but finite, fraction of each particle's orbit. The modification was used in order to speed up the calculation, which is otherwise extremely computationally expensive. We have performed several tests using a much smaller and more computationally tractable physical domain to study this. We find that when we use the same modification (constant field over a small portion of the orbit) we obtain the double-humped pitch angle distribution; but when we relax this assumption, the double-humped feature goes away. Thus, we conclude that the double-humped feature presented by GJ15 is not real.

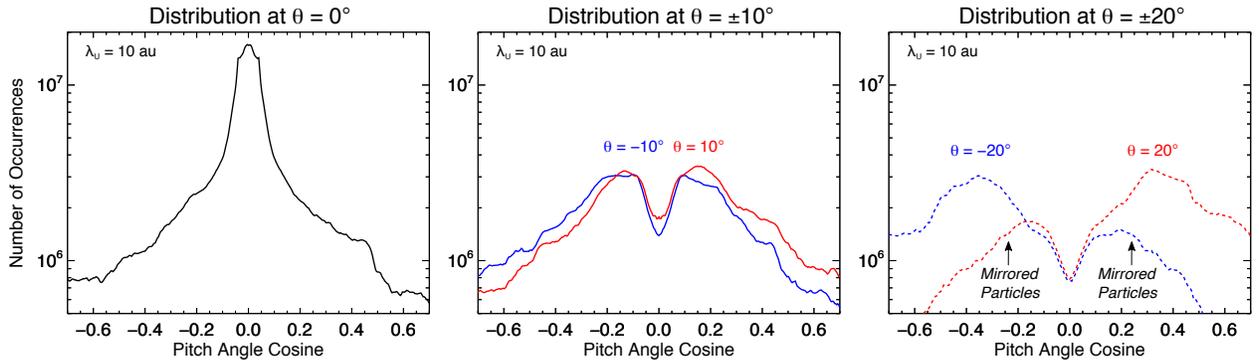

**Figure 8.** Particle pitch angle distribution in uniform mean ISMF ($\boldsymbol{B}_0$ directed towards +*z* axis) using the same turbulence field δ$\boldsymbol{B}$ from Figure 6 ($\lambda_U$ = 10 au, Realization #1, Case 1 from Table 1). The directions in the sky from which the distributions are extracted are shown in Figure 7.

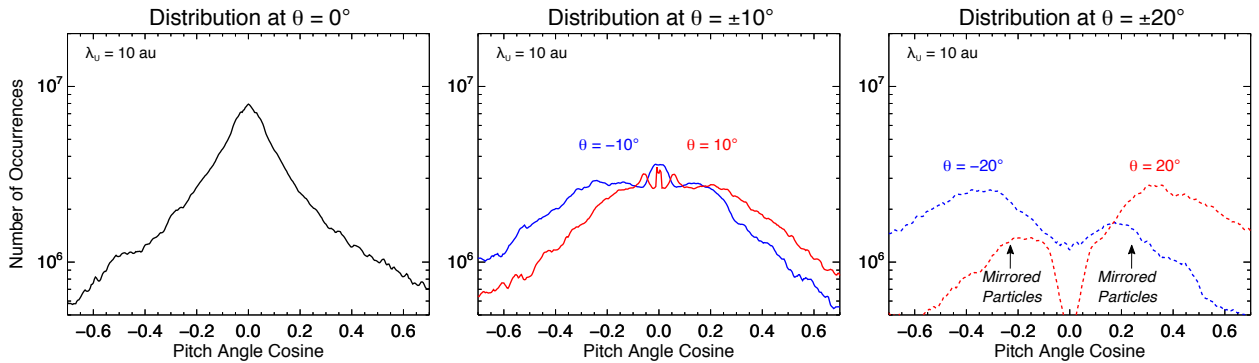

**Figure 9.** Same as Figure 7, except for different turbulence parameters ($\lambda_U$ = 500 au, Realization #1, Case 4 from Table 1).

At larger angles from $\boldsymbol{B}_0 \cdot \boldsymbol{r} = 0$ ($\theta = \pm 10°$), the pitch angle distributions become broader in both cases, and double-peaked for the $\lambda_U = 10$ au case, due to the mirroring of particles into backwards-propagating hemispheres. This is similar to the results presented by GJ15, supporting the importance of the magnetic mirror force's impact on the source particle distribution of the ribbon. For $\theta = \pm 20°$, the pitch angle distribution becomes less symmetric about pitch angle 90°, with more particles propagating in the forward hemisphere (away from $\boldsymbol{B}_0 \cdot \boldsymbol{r}$). This is because, at this larger angle from $\boldsymbol{B}_0 \cdot \boldsymbol{r} = 0$, fewer particles have mirrored due to their smaller initial pitch angles (or larger μ). The results for $\lambda_U = 10$ au and $\lambda_U = 500$ au are qualitatively similar, though the peaks in the pitch angle distributions are more broadly spread out for the $\lambda_U = 500$ au case due to enhanced mirroring from larger fluctuations in $\delta \boldsymbol{B}$.

## 4. Discussion and Conclusions

In this study we simulated the transport of PINS outside the heliopause in the presence of turbulence consistent with *Voyager 1* observations and a draped mean ISMF inferred from *IBEX* ENA observations. We tested the effects of the assumption of a uniform mean ISMF versus a draped mean ISMF derived from our MHD simulation of the heliosphere, as well as how different turbulence realizations and maximum fluctuation scales affect the ribbon fluxes at 1 au. In the following we discuss the implications of our results on the source of the ribbon and the turbulence properties in the VLISM.

### 4.1. Turbulence Fluctuation Scale Outside the Heliopause

The Kolmogorov-like, homogeneous turbulence that is observed by *Voyager 1* in the VLISM at scales <5 au is significantly stronger than the expected pristine turbulence from the interstellar medium (see Figure 2), suggesting it is not just of interstellar origin. We find that in order to reproduce large-scale structure of the *IBEX* ribbon (McComas et al. 2017), under the assumptions of our model, the turbulence observed by *Voyager* 1 should remain limited to length-scales smaller than approximately $\lambda_U \sim 10$ au. We find that the cases where $\lambda_U < 100$ au, and in particular the solution for $\lambda_U \sim 10$, produce the most realistic solutions. This turbulence could either be a result of the interaction between the heliosphere and the pristine interstellar turbulence, or it could be entirely of heliospheric origin. For example, Zank et al. (2017) showed that the largely compressive fluctuations observed by *Voyager 1* from ~2013.4 to 2014.6 (Burlaga et al. 2015) are consistent with fast-mode waves propagating outside the heliopause, originating from fast and slow-mode waves that were refracted at the heliopause and propagated outside the heliopause at highly oblique angles to the ISMF. However, subsequent observations by *Voyager 1* showed that from ~2015.4 to 2016.7 the fluctuations were dominantly transverse to the mean field and not compressive, though the uncertainty of this result was large (Burlaga et al. 2018; see also Fraternale et al. 2019). As suggested by Burlaga et al. (2018), these observations can be verified by future *Voyager 1* and *2* observations of the VLISM turbulence.

Nevertheless, Zank et al. (2017) predict that locally-generated turbulence from fast-mode waves emanating from the heliosphere is likely superposed on the pristine interstellar turbulence spectrum. The dependence of this superposed turbulence spectrum on distance from the heliosphere, however, is not well understood, which is the reason why in our study we assumed that the turbulence power spectrum was the same everywhere outside the heliopause. While this assumption can be analyzed in more detail in a future study, we believe our results are robust since the spatial distribution of the ribbon source is largely concentrated within a few tens of au from the heliopause (e.g., Zirnstein et al. 2019) where, presumably, the turbulence observed by *Voyager 1* is strongest.

### 4.2. Ribbon Structure: Single or Double Peak?

The results presented in this study show that the double-peaked ribbon feature predicted by GJ15 does not exist, at least not for the particular properties of turbulence presented (circularly polarized, isotropic turbulence). The reason for this was discussed above in Section 3.3. Instead, the ribbon is single-peaked, but still slightly broadened due to the magnetic mirroring of particles in turbulence. However, the broadness of the ribbon is a combination of (1) the draping of the ISMF around the heliopause and (2) the magnetic mirroring of particles in turbulence. This is apparent by comparing the simulation results between a uniform mean ISMF and with a draped mean ISMF with and without turbulence (Figure 3).

Through our analysis of the discrepancy in the results presented here and those presented in GJ15 regarding to the existence of the double-humped feature, we have found that the shape of the ribbon ENA emission profile near 90° pitch angle can vary significantly depending on the nature of the turbulence. It remains an open question as to whether a double-peaked ribbon might arise for different turbulence properties. Under the constraint that the normalization of the turbulence power is consistent with *Voyager 1* observations (Figure 2), and assuming that the turbulence is isotropic and Kolmogorov-like, we doubt that a double-peaked ribbon is likely, especially in a draped mean ISMF which would smear out any double-peaked structure.

### 4.3. Fine Structure in the Ribbon

While the existence of the double-peaked ribbon appears unlikely based on the results presented in this study, it may be possible that smaller fine structure exists in the ribbon, like that possibly observed in the first *IBEX* maps (McComas et al. 2009b). Our simulation results (e.g., Figure 6) suggest that neighboring pixels on the order of a few degrees may differ in intensity due to particle interactions with turbulence. Moreover, evolution of these intensities over time can be expected. The time over which ENA fluxes evolve depends on the turnover time of turbulent eddies at the scales over which we are observing.

For example, at the *IBEX* angular resolution of 6° (Funsten et al. 2009), and with a ribbon source distance of approximately 150 au from the Sun (Swaczyna et al. 2016a; Zirnstein et al. 2018, 2019), the arc distance of the ribbon's source with this opening angle corresponds to 16 au. The turnover time for turbulence with wavelength $\lambda = 16$ au is approximately proportional to the ratio of wavelength to the Alfvén speed outside the heliopause, $\lambda/v_A$. In a 4 µG magnetic field and plasma density of 0.1 $cm^{-3}$, $v_A = 28$ km $s^{-1}$. Thus, the turnover time is on the order of $\lambda/v_A =$ 16 au/28 km $s^{-1} \cong 2.7$ yr. However, small changes in turbulence, and thus small changes in particle pitch angle, may significantly affect the ENA flux observed at 1 au by *IBEX*. Therefore, the expected time over which fine structure in the ribbon changes is likely smaller than this. Suppose the turbulent eddy turnover time results in, on average, a change in phase of 90°. A small change in phase ($\lesssim 10°$) could change the particle pitch angles significantly enough such that the expected time scale over which fine structure in the *IBEX* ribbon changes could possibly occur over a few months or less.

Whether *IBEX* has sufficient measurement statistics to observe these small changes is currently not well understood. However, the *Interstellar Mapping and Acceleration Probe* (*IMAP*) will have better angular resolution compared to *IBEX*, between 4° and 2° for *IMAP-Hi* and *IMAP-Ultra*, respectively, and better statistics (McComas et al. 2018a). With these improvements, *IMAP* could potentially observe even smaller structure in the ribbon and fluctuations in the fine structure corresponding to the evolution of smaller turbulent eddies.

*Acknowledgments.* E.Z., M.D., and J.H. acknowledge support from NASA grant 80NSSC17K0597. This work was also partly funded by the *IBEX* mission as a part of the NASA Explorer Program (80NSSC18K0237). J.G. acknowledges NASA grant NNX15AJ71G and useful discussions on this topic with J.R. Jokipii. R.K. acknowledges support from the Max-Planck/Princeton Center for Plasma Physics and NSF Grant No. AST-1517638. E.Z. and J.G. acknowledge helpful discussions at the team meeting "The Physics of the Very Local Interstellar Medium and Its Interaction with the Heliosphere" supported by the International Space Science Institute in Bern, Switzerland. E.Z. thanks Len Burlaga for providing the power spectral density data from *Voyager 1*. The work reported in this paper was partly performed at the TIGRESS high performance computer center at Princeton University which is jointly supported by the Princeton Institute for Computational Science and Engineering and the Princeton University Office of Information Technology's Research Computing department.